\documentclass[letterpaper, preprint, paper,11pt]{AAS}	

\usepackage{bm}
\usepackage{amsmath}
\usepackage{subfigure}
\usepackage[colorlinks=true, pdfstartview=FitV, linkcolor=black, citecolor= black, urlcolor= black]{hyperref}
\usepackage{overcite}
\usepackage{footnpag}			      	
\usepackage{graphicx}
\usepackage{listings}
\usepackage{footnote}
\makesavenoteenv{tabular}
\makesavenoteenv{table}

\usepackage{url}
\makeatletter
\g@addto@macro{\UrlBreaks}{\UrlOrds}
\makeatother

\usepackage{tikz}
\usetikzlibrary{shapes,arrows}
\tikzstyle{block} = [draw, fill=blue!20, rectangle, 
minimum height=3em, minimum width=6em]

\tikzstyle{sum} = [draw, fill=blue!20, circle]
\tikzstyle{input} = [coordinate]
\tikzstyle{output} = [coordinate]
\tikzstyle{pinstyle} = [pin edge={to-,thin,black}]

\PaperNumber{XX-XXX}

\begin{document}

\title{Space Navigator: a Tool for the Optimization of Collision Avoidance Maneuvers}

\author{Leonid Gremyachikh\footnotemark \footnotetext{National Research University Higher School of Economics, Laboratory of Methods for Big Data Analysis, Myasnitskaya St. 20, 101000 Moscow, Russia.},
Dmitrii Dubov\footnotemark[1],
Nikita Kazeev\footnotemark[1] \footnotemark \footnotetext{Yandex School of Data Analysis, Timura Frunze St. 11/2, 119021 Moscow, Russia.},
Andrey Kulibaba\footnotemark \footnotetext{JSC "Russian Space Systems", Aviamotornaya str., 53, 111024, Moscow, Russia.},
Andrey Skuratov\footnotemark \footnotetext{Phygitalism, Presnensky Val St. 27/8, 123557 Moscow, Russia.},
Anton Tereshkin\footnotemark[4], 
Andrey Ustyuzhanin\footnotemark[1] \footnotemark[2],
Lubov Shiryaeva\footnotemark[3] and
Sergej Shishkin\footnotemark[3]
}

\maketitle{} 		

\begin{abstract}
    The number of space objects will grow several times in a few years due to the planned launches of constellations of thousands microsatellites. It leads to a significant increase in the threat of satellite collisions. Spacecraft must undertake collision avoidance maneuvers to mitigate the risk. According to publicly available information, conjunction events are now manually handled by operators on the Earth. The manual maneuver planning requires qualified personnel and will be impractical for constellations of thousands satellites. In this paper we propose a new modular autonomous collision avoidance system called "Space Navigator". It is based on a novel maneuver optimization approach that combines domain knowledge with Reinforcement Learning methods.
\end{abstract}

\section{Introduction}

    It is estimated that there are about 22,000 pieces of debris, measuring at least 10 cm in diameter, and over 600,000 pieces larger than 1 cm.\cite{n_debris} All of them travel fast enough to damage a spacecraft. Currently, there are about 1,800 operational satellites orbiting the Earth.\footnote[5]{\url{http://m.esa.int/Our_Activities/Operations/Space_Debris/Space_debris_by_the_numbers}} With such number of objects, satellite collision avoidance maneuvers (CAM) are necessary, for example, Landsat 7 executed 4 maneuvers in 2017.\footnote[6]{\url{https://satellitesafety.gsfc.nasa.gov/maneuvers.html}} At the same time the amount of working satellites is increasing. For example, SpaceX is planning to launch 4,425 units by 2024.\cite{SpaceX}  
    
    The other thing to consider is that if two large items collide in space, the result is a huge number of new dangerous objects. \cite{Space_Debris_Threat}  Only Iridium-Cosmos collision in 2009 produced almost 1,850 pieces of debris larger than 10 cm and thousands more smaller pieces. \cite{cos_iri}

    According to publicly available information, conjunction events are now manually handled by operators on the Earth.\cite{esa_process} The manual decision-making process requires qualified personnel and will be impractical for constellations of thousands satellites. Therefore, there is a need to develop an automated collision avoidance system. ESA has recently launched a tender for such a system. \footnote[8]{\url{https://artes.esa.int/funding/autonomous-collision-avoidance-system-ngso-artes-3a093}}
    
    Designing an automated collision avoidance system is a challenging task. The optimal maneuver must balance multiple factors such as collision probability, propellant consumption, mission objective, and radio visibility zones. In the face of the increasing number of space objects, it might be necessary to take account several debris pieces when planning maneuvers.\cite{multi-debris, cluster}
    
    There are several systems for CAM calculation such as CORAM (Reference~\citenum{coram}) and OCCAM (Reference~\citenum{occam}).
    
    CORAM is employed in Collision Avoidance service by ESA's Space Debris Office and provides an operator with comprehensive information about conjunctions, maneuvers, and trajectories. This system provides the capacity to cope with Multi-Encounter and Multi-Maneuver cases. CORAM allows to define a flexible optimization function, taking into account the collision probability, maneuver size, and miss distance.
    
    OCCAM is a software tool for fast CAM computation based on analytical formulation of the collision problem.\cite{approx} This system is able to cope with conjunction with just one dangerous space object. OCCAM supports only three optimization goals: conditional optimization of fuel consumption, collision probability minimization, and miss distance maximization. Among the advantages, there is computation speed and a variety of methods of collision probability estimation. The demo version of OCCAM is freely available online.\footnote{\url{http://sdg.aero.upm.es/index.php/online-apps/occam-lite}}

    In this paper, we propose a new system, "Space Navigator" (SpaceNav), based on a novel maneuver optimization algorithm that combines domain knowledge with Reinforcement Learning (RL) methods. We also describe the SpaceNav architecture and system features. One of the key features of SpaceNav is modularity which allows the system to be used by different satellite operators with different propagators, methods of collision probability estimation, and optimization requirements. Furthermore, SpaceNav can be configured to cope with various tasks such as Multi-Encounter, Multi-Maneuver, and Maneuvering in a Cluster.
    
    SpaceNav is a result of joint efforts of the Roscosmos Corporate Academy and the Laboratory of Methods for Big Data Analysis at the NRU Higher School of Economics. Virtual Reality interface is a contribution of Phygitalism.
   
    This paper is organized as follows. First, we briefly introduce the concept of Reinforcement Learning. Next, we present the SpaceNav architecture and discuss its capabilities. Then we turn to the maneuver optimization algorithms and introduce a novel application of a well-established Reinforcement Learning algorithm for the purpose. Finally, we discuss the experimental sample of dangerous situations and show the optimization algorithms performance.

\section{Reinforcement Learning}

    Reinforcement Learning (RL) is a field of Machine Learning. The key idea of RL is to develop Agent which provides optimal Actions for some State of some Environment so as to maximize a numerical Reward signal.\cite{RL_intro} This approach has already been applied to the spacecraft maneuvering task. \cite{rl_space1, rl_space2}
 
    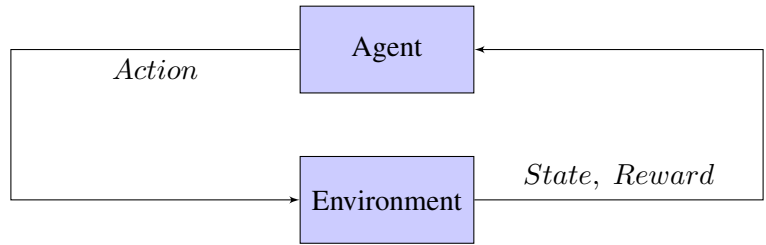
\begin{figure}[!htb]
        \centering
        \begin{tikzpicture}[auto, node distance=2cm,>=latex']
            \node [block] (agent) {Agent};
            \node [block, below of=agent] (env) {Environment};
            \draw [draw,->] (agent) -- node {$Action$} +(-5,0) |- (env);
            \draw [draw,->] (env) -- node {$State,\;Reward$} +(5,0) |-  (agent);
        \end{tikzpicture} 
        \caption{The Reinforcement Learning scheme}
        \label{fig:rl}
    \end{figure}
    
    The simplified training process of Agent is presented in Figure \ref{fig:rl}. State of Environment represents the current information about objects position, velocities, and any additional parameters. State get to the input of Agent which provides Actions. After that, Environment implements the Actions and State is changed to another State. Also, Environment returns an assessment (Reward) of Action, new State, or whole session. Reward helps to improve the Agent.
    
    Such a principle of training has benefits.\cite{RL_intro}
    Reinforcement Learning, being a numeric method, does not require the Environment to be simple enough to allow for an explicit mathematical solution for the problem of finding the best action. In contrast with numerical optimization methods, RL allows building an operator that explicitly maps State into the optimal Action (usually a neural network), without requiring the reward function to be differentiable.

\section{Space Navigator}
    \subsection{About}
        Operator decisions depend on a variety of mission-specific concerns. The state-of-the-art systems mentioned in the introduction are designed to provide the operator with full information about the conjunction event and possible avoidance maneuvers as input for decision-making. In contrast, Space Navigator is built from the ground up as an automated modular system, which is loaded by a set of evaluations important to a particular case. The optimization objective function can be arbitrarily defined. This allows taking into account not only collision probability and fuel consumption, but also other complicated optimization requirements, for example, allowed orbit deviation. SpaceNav allows simultaneously considering many space objects (up to 10 in experiments). 
        
    \subsection{Architecture}
            
\begin{figure}[!htb]
            \centering
            \begin{tikzpicture}[auto, node distance=5cm,>=latex']
                \node [input, name=pos,] {};
                \node [input, name=req, below of=pos, node distance=1cm] {}; 
                \node [block, right of=pos] (spacenav) {SpaceNav};
                \node [output, right of=spacenav] (output) {};
                \draw [draw,->] (pos) -- node {$objects \; pos.$} (spacenav);
                \draw [draw,->] (req) -- node {$requirements$} +(4,0) -|  (spacenav);
                \draw [->] (spacenav) -- node {$maneuver$}(output);
            \end{tikzpicture}
            \caption{SpaceNav pipeline}
            \label{fig:spacenav_pipeline}
        \end{figure}
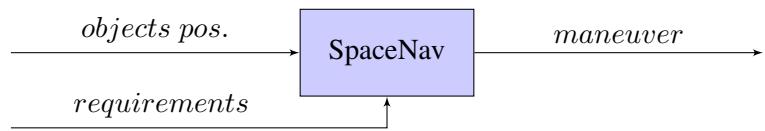

        The SpaceNav pipeline is presented in Figure \ref{fig:spacenav_pipeline}. Inputs for SpaceNav are:
        
        \begin{itemize}
            \item Space objects coordinates, velocities, their uncertainties, and epoch;
            \item Optimization requirements such as a threshold of collision probability or miss distance, thresholds of orbital elements deviation, maneuver restrictions, maximum fuel consumption, fuel level and so on.
        \end{itemize}

\begin{figure}[!htb]
            \centering
            \begin{tikzpicture}[auto, node distance=3cm,>=latex']
                \node [input, name=pos,] {};
                \node [input, name=req, below of=pos, node distance=1cm] {}; 
                \node [sum, right of=pos, node distance=4cm] (concat) {};
                \node [block, right of=concat, ] (agent) {Agent};
                \node [sum, right of=agent] (act2man) {};
                \node [output, right of=act2man, node distance=4cm] (man) {};
                \node [output, below of=man, node distance=1cm] (info) {};
                
                \node [block, below of=agent, minimum width=8em] (simulator) {Env. Simulator};
                
                \draw [draw,->] (pos) -- node {$objects \; pos.$} (concat);
                \draw [draw,->] (req) -- node {$requirements$} +(4,0) -|  (concat);
                \draw [draw,->] (concat) -- node [name=state] {$state$} (agent);
                \draw [draw,->] (agent) -- node [name=action] {$action$} (act2man);
                \draw [draw,->] (act2man) -- node {$maneuver$} (man);
                
                \draw [->] (action) |- (simulator);
                \draw [->] (simulator) -- node {$reward$} (agent);
                \draw [->] (state) |- (simulator);
                
                \draw [color=gray,thick](3.5,-4) rectangle (10.5,1);
    	        \node at (3.5,1) [above=5mm, right=0mm] {\text{SpaceNav}};
            \end{tikzpicture} 
            \caption{The simplified architecture of SpaceNav}
            \label{fig:spacenav_arch}
        \end{figure}
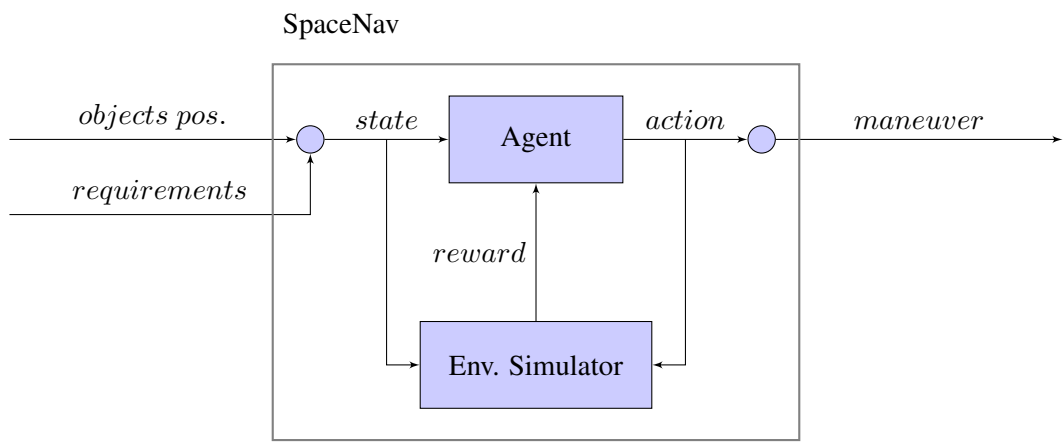

        SpaceNav returns one or several maneuvers according to the maximum value of an optimized objective function. Having a convenient API, SpaceNav allows easily utilizing different propagators, optimization strategies, and collision probability computation approaches. The simplified architecture of SpaceNav is shown in Figure \ref{fig:spacenav_arch}.

         The Agent module gets optimization requirements and a description of a dangerous situation, which is a situation with critical conjunctions with one or more objects. Such an input represents State. After calculation, Agent provides a maneuver, i. e. Action, for the input State.
        
        The Agent module could represent an optimization instrument, such as Grid Search or Stochastic Optimization model. Such a model solves the optimization problem for each collision situation without using any previous experience. Therefore, it could be a general-purpose module which require a lot of sessions. On the other hand, Agent could represent some trained specialized model, such as Neural Network. Such a model could rapidly provide CAM, but it would not necessary be be a global optimum. Eventually, Agent could represent several models at once.
        
        According to the selected model, the Agent is trained. For training, it is necessary to gain some response from the Environment and somehow evaluate the obtained Actions (maneuvers) with Reward. In our case, Environment represents a simulator with propagator of orbital movement. Simulator returns the values of the optimized parameters at the end of the propagation session. The goal is to choose the best maneuver according to the maximum value of Reward. Depending on the model, the training process could be carried out for a specific set of input parameters or in advance.
        
        The next question is how to evaluate Agent Actions. SpaceNav provides flexible and easy-to-tune Total Reward function for different purposes. For example, a user could set critical values of collision probability, fuel consumption, and semi-major axis deviation as thresholds. This approach will help to obtain optimal maneuvers also considering maintenance of semi-major axis. Users can also adjust the Total Reward function for their own purpose or replace it with a custom one.

        \begin{figure}[htb]
        	\begin{center}
        		\includegraphics[width=0.7\textwidth,]{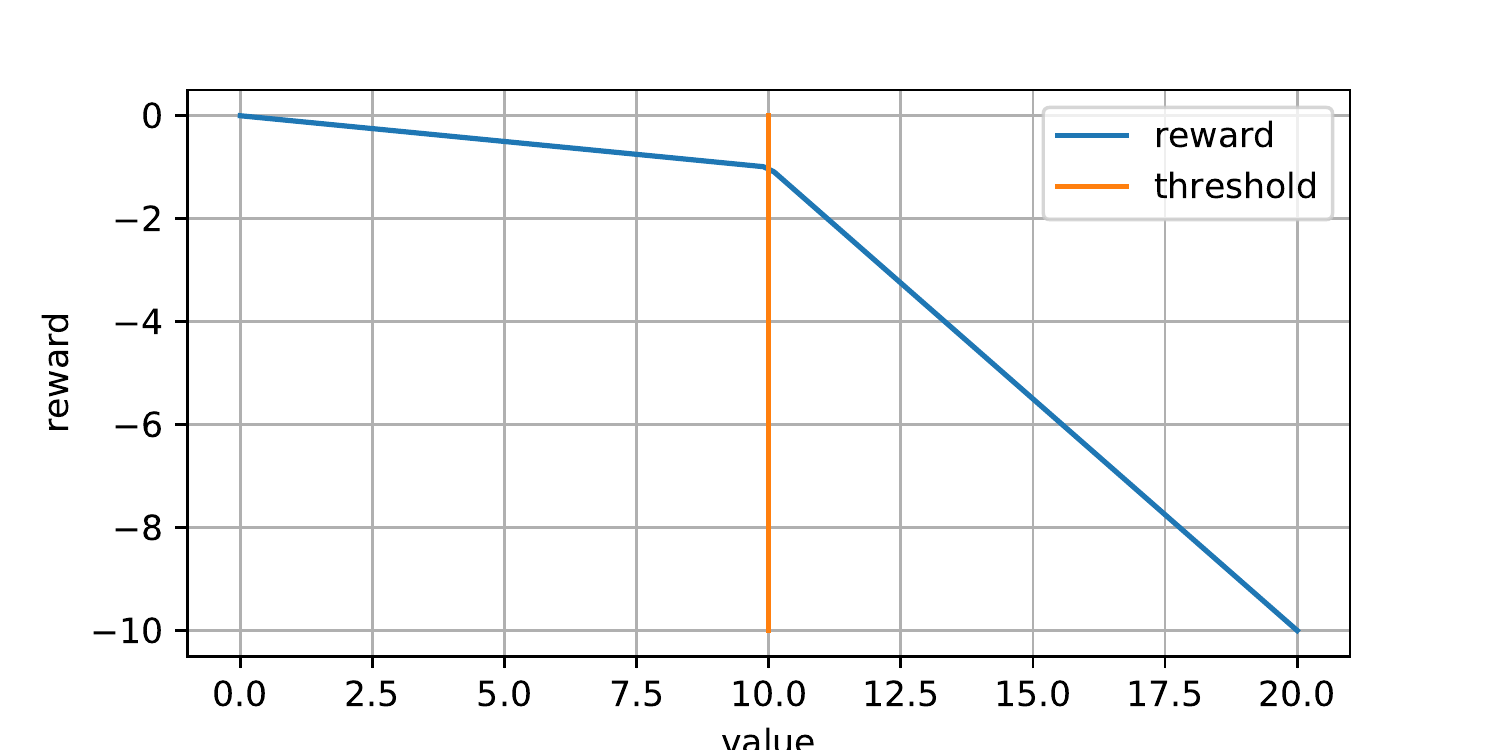}
        	\end{center}
        	\caption{The reward component function, threshold = 10}
        	\label{fig:reward_component}
        \end{figure}
        
        In this paper, we provide the function of Total Reward which is a sum of several reward components, such as a penalty for the collision probability, fuel consumption, and trajectory deviation (Equation~\eqref{eq:total_reward}). Each reward component is a piecewise linear function of the component value and threshold. This function consists of two linear areas. The first one, with a low slope, is for component values which are lower than the threshold and the second one, with a high slope, is for component values which exceed the threshold (Figure \ref{fig:reward_component}). The Total Reward function is designed with the purpose of dramatically increasing the penalty if the values are above the threshold.
        
        \begin{equation}
        	\label{eq:total_reward}
        	R_{total} 
        	= R_p (P_{collision}) + R_{dV} (dV_{maneuver}) + \dots 
        	= \sum_i R_i (component_i)
        \end{equation}
        

    
    \subsection{Visualization}
        
        We also developed an experimental Virtual Reality (VR) frontend to aid with outreach activities. The VR system includes:
        
        \begin{itemize}
            \item a hologram of the Earth and space objects with orbital tracks;
            \item a 2-D interface with a scaled visualization of the maneuvering process;
            \item an ability to interactively select the evasion trajectory;
            \item a collision alert signal;
            \item a collision animation.
        \end{itemize}
 
\subsection{Simulation tuning}
SpaceNav relies on simulation of space objects motion. To be precise, the simulation must take into account the variability of atmospheric, solar, and geomagnetic conditions. Moreover, their influence depends on the properties of a space object, that may not be known for debris objects.\cite{LEO_lifetime} However, since those parameters influence the object motion, it might be possible to recover them by observing the object motion.

The problem can be formulated as follows: given the history of a space object motion and a simulator state, find the simulator parameters that would provide the best match between the simulated and real space object motion. We think that Reinforcement Learning can be applicable here as well. We did a pilot study on simulator tuning which produced the promising result.\cite{simulation_tuning} As an ad-hock feasibility test of this approach for space motion simulation, we used the open source poliastro simulator which supports modelling of atmospheric drag. \cite{poliastro} Using the Cross Entropy method we were able to tune the simulator to find the unknown space object cross-sectional area.

\section{Algorithms}

        For the sake of brevity, we use the following notations to indicate the direction of maneuver:
        
        \begin{itemize}
            \item in-track maneuver -- a maneuver collinear to the satellite's velocity vector;
            \item in-plane maneuver -- a maneuver in the satellite's orbital plane;
            \item out-of-plane maneuver -- a maneuver in any direction.
        \end{itemize}

        In this paper we describe Cross Entropy (CE) and In-track Grid Search (GS) methods.

        The GS method provides in-track maneuver before $n + 0.5$ orbital periods of a dangerous conjunction. This model iterates over the grid of in-track maneuvers in the range from $-dV_{max}$ to $dV_{max}$, where $dV_{max}$ is the maximum allowed fuel consumption. We developed two modifications of GS:
        \begin{itemize}
            \item Baseline mode: takes into account only closest dangerous object. After each maneuver, the algorithm is restarted and offers new maneuvers if necessary.
            \item General mode: takes into account all input objects at once and provide only one maneuver.
        \end{itemize}
        
        CE is a method based on the Stochastic Optimization approach.\cite{stoch_opt}. This method is able to offer maneuvers in different directions and find the optimal maneuver epoch. The first step is to choose an initial maneuver and an appropriate random distribution. The expected value $E$ of the distribution is equal to initial maneuver parameters. The distribution will be used for generating new maneuvers based on the initial one. Next, the algorithm repeats following iterations:

        \begin{enumerate}
            \item generate a random sample of maneuvers from the distribution;
            \item evaluate each maneuver by a reward;
            \item select some maneuvers with the best reward;
            \item shift $E$ in the direction of the selected maneuvers;
            \item additional modifications, such as a dispersion decay.
        \end{enumerate}
        
        Iterations are repeated until the stopping criterion, such as a limit on the number of iterations, is satisfied. CE is a well-known method and there are many ways to improve the algorithm, for example, by an introduction of a learning rate and a dynamic decay of the standard deviation of the maneuver parameters during sampling.\cite{CE_features}
        
        Stochastic Optimization approach is explored in the literature on maneuvers optimization.\cite{multi-debris} However, we have not seen the use of CE in this field. In addition to the immediate maneuvers optimization, the CE method can be effectively used for tuning maneuvers obtained by other models or theoretically.
        
        We compare the following algorithms (designation of the algorithm is mentioned in the brackets):

        \begin{itemize}
            \item In-track Grid Search:
            \begin{itemize}
                \item Baseline mode (baseline);
                \item General mode (GS);
                \item General mode with CE in-plane tuning (GS + CE).
            \end{itemize}
            \item Cross-Entropy method:
            \begin{itemize}
                \item in-track, maneuver half an orbital period before the conjunction (CE in-track half);
                \item in-plane, maneuver half an orbital period before the conjunction (CE in-plane half);
                \item out-of-plane, maneuver half an orbital period before the conjunction (CE out-of-plane half);
                \item in-track, automatic maneuver timing (CE in-track auto);
                \item in-plane, automatic maneuver timing (CE in-plane auto);
                \item out-of-plane, automatic maneuver timing (CE out-of-plane auto).
            \end{itemize}
        \end{itemize}        
        
\section{Results}
    \subsection{Experiment description}
    
        For the experiment, we assumed the worst-case scenario. According to the assumption, a collision warning is one orbital period before a dangerous conjunction. After the first encounter, there are nine other dangerous objects, whose trajectories almost intersect the trajectory of the protected object. Such additional objects represent obstacles for maneuvers because taking into account only one object will lead to a potential collision with another object. The duration of each simulation is twenty-four hours plus one orbital period of the protected object. The time before the dangerous conjunction with the first object is one orbital period. The epochs of the other dangerous conjunctions are randomly located on the simulated time interval. We used the collision probability computation method proposed in (Reference \cite{chenbai}) and a Keplerian propagator (Reference \cite{pykep}). To improve the stochastic optimization results, CE-based algorithms are run two times for each situation and the best result is recorded.

        To evaluate and train the Agent models, we have developed a dangerous situations generator. The sizes of space objects, angles of intersection of orbits, and other parameters are randomly generated from distributions described in Appendix A. Using the generator we have obtained a sample of 100 random situations. An example of a generated situation could be seen in Appendix B.

        \begin{table}[htbp]
            \fontsize{10}{10}\selectfont
            \caption{Thresholds}
            \label{tab:thr}
            \centering 
            \begin{tabular}{l|l}
                \hline
                parameter                                 & threshold \\ \hline
                collision probability                     & 1e-4      \\ 
                $a$ - semi-major axis deviation (meters)    & 200       \\ 
                $e$ - eccentricity deviation                & 0.01      \\ 
                $i$ - inclination deviation (rad)            & 0.01      \\ 
                $\Omega$ - longitude of the ascending node (rad) & 0.01      \\ 
                $\omega$ - argument of periapsis (rad)           & 0.01      \\ 
                fuel ($m^2 / s$)                          & 1.0       \\ \hline
                \end{tabular}
        \end{table}
        
        Table \ref{tab:thr} shows the threshold values used for the experiment. The algorithms were required not only to mitigate the collisions risk  but also to remain within the specified limits of the trajectory deviation.
    
    \subsection{Evaluation Results}
    
        Table \ref{tab:results} shows the results of evaluation of the performance of the algorithms on the 100 randomly generated dangerous situations. Also in the appendices we provide detailed results for one of the generated dangerous situations. The situation description is in Appendix B, the obtained maneuvers and result values are in Appendix C, and conjunctions tables are in Appendix D.

\begin{table}[htbp]
\fontsize{9}{9}\selectfont
\caption{Results (\%), where: 
top $10\%$ -- model reward differs from the best model by no more than 10 percent,
$\leq$ thr -- all values are below the thresholds,
o/c baseline -- model overcomes baseline in terms of reward,
o/c GS -- model overcomes GS in terms of reward,
$P_{c}$ -- total collision probability.
}
\label{tab:results}
\centering 
\begin{tabular}{l|l|l|l|l|l|l|l}
\hline
                  & top $10\%$ & $\leq$ thr & o/c baseline & o/c GS & $P_{c} \leq$ 1e-4 & $P_{c} \leq$ 2e-4 & $P_{c} \leq$ 1e-3 \\ \hline
baseline          & 0          & 20         & -            & 31     & 70                & 89                & 97                \\ 
GS                & 3          & 23         & 83           & -      & 68                & 84                & 97                \\ 
GS+CE             & 53         & 66         & 100          & 100    & 99                & 100               & 100               \\ \hline
CE in-track half  & 3          & 18         & 57           & 20     & 66                & 90                & 93                \\ 
CE in-plane half  & 47         & 46         & 91           & 88     & 96                & 100               & 100               \\ 
CE out-plane half & 48         & 54         & 86           & 80     & 99                & 99                & 100               \\ \hline
CE in-track auto  & 64         & 55         & 94           & 96     & 91                & 98                & 100               \\ 
CE in-plane auto  & 66         & 57         & 95           & 92     & 94                & 98                & 100               \\ 
CE out-plane auto & 72         & 68         & 97           & 96     & 96                & 99                & 100               \\ \hline
\end{tabular}
\end{table}

The results show that in the majority (68\%) of cases SpaceNav is able to find maneuvers for these worst case scenario satisfying all the complex constraints. In almost every case (99\%) it reduces the total collision probability to the level of $2 \cdot 10^{-4}$.

The reward function in this example has been configured to aggressively save fuel and maintain orbit deviation small. By configuring the reward function it is possible to achieve any desired level of collision risk. 
        
The results provide useful insight into the behaviour of the optimization algorithms. CE with automatic maneuver timing outperforms CE with fixed timing -- in the case of multiple conjunctions optimal maneuver time is not half an orbital period before the first conjunction, and CE seems to find it. Performance of CE with zero initialization (CE in-plane half) and CE initialized with the Grid Search are close with GS+CE slightly better. CE is a stochastic algorithm, so a good initial approximation allows it make better use of the limited number of iterations.
        
\section{Roadmap}
    A prototype of the SpaceNav project is completed. Further roadmap includes the following items:
    
    \begin{enumerate}
        \item add fast initial maneuver approximation using Neural Networks;
        \item develop GUI for SpaceNav;
        \item add optimization of sequence of maneuvers;
        \item integration with data sources, such as DISCOS;
        \item elaboration of integration into systems of the ground control complex of space objects.
	\end{enumerate}
	
    We also experiment with various other models such as Neural Networks\cite{thesis_dmitriy}, Evolution Strategies\cite{thesis_dmitriy}, and Monte Carlo Tree Search\cite{thesis_leonid}, the results of which are not provided in this paper.
    
\section{Conclusion}
    In this paper, we present an autonomous modular collision avoidance system called SpaceNav. This system is based on the Reinforcement Learning approach to for maneuver optimization. Furthermore, we provide a description of the new maneuver optimization algorithm and the objective function (reward function). Also, we show the results of experimental evaluation of SpaceNav on a sample of 100 randomly generated dangerous situations, as well as full information of one particular situation.

\section{Acknowledgment}
    We thank Fedor Ratnikov for his useful criticism and fruitful discussions.

The research was carried out with the financial support of the Ministry of Science and Higher Education of the Russian Federation within the framework of the Federal Target Program “Research and Development in Priority Areas of the Development of the Scientific and Technological Complex of Russia for 2014-2020”. Unique identifier -- RFMEFI58117X0023.

\appendix
\clearpage

\section*{Appendix A: generator distributions}
    
    Distribution of protected object parameters:
    
    \begin{itemize}
        \item semi-major axis (m): $a \sim \mathcal{U}(7\cdot10^6, 8\cdot10^6)$;\footnote{\url{https://upload.wikimedia.org/wikipedia/commons/b/b4/Comparison_satellite_navigation_orbits.svg}}
        \item eccentricity: $e \sim \mathcal{U}(0 , 0.003)$;
        \item inclination (rad): $i \sim \mathcal{U}( 0, 2\pi)$;
        \item longitude of the ascending node (rad): $\Omega \sim \mathcal{U}( 0, 2\pi)$;
        \item argument of periapsis (rad): $\omega \sim \mathcal{U}( 0, 2\pi)$;          
        \item mean anomaly (rad): $v \sim \mathcal{U}( 0, 2\pi)$;
        \item radius (m): $r \sim \mathcal{U}(0.3,\,55)$.\footnote{\url{http://www.businessinsider.com/size-of-most-famous-satellites-2015-10}}
    \end{itemize}

    Distribution of debris object parameters:
    
    \begin{itemize}
        \item angle between protected and debris orbital planes (rad): $\alpha \sim \mathcal{U}(0.5,2.64)$;
        \item position at the conjunction moment:
        \begin{itemize}
            \item first conjunction: $x_{debris} \sim \mathcal{N}(x_{protected}, 50)$, same for $y$ and $z$ (m);
            \item other conjunctions: $x_{debris} \sim \mathcal{N}(x_{protected}, 500)$, same for $y$ and $z$ (m).
        \end{itemize}
        \item velocity vector magnitude (lies on a debris object plane and is tangent to the Earth): $v_{debris} \sim \pm \; \mathcal{N}(v_{protected}, 0.05)$ (m/s);
        \item radius (m) $r \sim \mathcal{U}(0.05,\,1)$.\footnote{\url{https://m.esa.int/Our_Activities/Operations/Space_Debris/Space_debris_by_the_numbers}}
    \end{itemize}

\clearpage

\section*{Appendix B: an example generated dangerous situation}\label{app:description}
    Tables \ref{tab:env1} and \ref{tab:env2} present one danger situation from the generated sample. Simulation interval - from 6599.921 to 6601.0 (mjd2000).

\begin{table}[htbp]
\fontsize{9}{9}\selectfont
\caption{Dangerous situation - part 1}
\label{tab:env1}
\centering 
\begin{tabular}{l|r|r|r|r|r|r}
\hline
                          & PROTECTED   & DEBRIS0     & DEBRIS1     & DEBRIS2     & DEBRIS3     & DEBRIS4     \\ \hline
$a$       & 7530537.215 & 7360115.107 & 8033345.687 & 7682829.226 & 6203113.774 & 7679322.768 \\ 
$e$              & 0.003       & 0.025       & 0.060       & 0.022       & 0.212       & 0.017       \\ 
$i$        & 0.562       & 1.555       & 0.896       & 1.146       & 2.103       & 0.591       \\ 
$\Omega$          & 2.551       & 1.809       & 5.957       & 2.022       & 3.738       & 0.852       \\ 
$\omega$        & 0.153       & 5.915       & 6.143       & 5.930       & 3.677       & 5.567       \\ 
$v$       & 2.153       & 3.103       & -0.000      & -0.076      & -3.139      & -0.068      \\ \hline
epoch & 6600.000    & 6600.000    & 6600.389    & 6600.791    & 6600.887    & 6600.923    \\ \hline
$r$                    & 20.686      & 0.738       & 0.367       & 0.562       & 0.564       & 0.276       \\ \hline
\end{tabular}
\end{table}

\begin{table}[htbp]
\fontsize{9}{9}\selectfont
\caption{Dangerous situation - part 2}
\label{tab:env2}
\centering 
\begin{tabular}{l|r|r|r|r|r}
\hline
                          & DEBRIS5     & DEBRIS6     & DEBRIS7     & DEBRIS8     & DEBRIS9     \\ \hline
$a$       & 8738088.965 & 8101742.447 & 7084346.824 & 7150262.637 & 7468758.271 \\ 
$e$              & 0.140       & 0.072       & 0.063       & 0.056       & 0.006       \\ 
$i$        & 0.376       & 1.856       & 2.168       & 2.882       & 2.484       \\ 
$\Omega$          & 0.657       & 3.769       & 4.917       & 3.179       & 2.870       \\ 
$\omega$        & 2.312       & 0.527       & 3.765       & 0.461       & 3.314       \\ 
$v$       & 0.005       & -0.002      & -3.105      & 3.140       & -3.134      \\ \hline
epoch & 6600.581    & 6600.061    & 6600.597    & 6600.238    & 6600.652    \\ \hline
$r$                    & 0.107       & 0.053       & 0.320       & 0.674       & 0.895       \\ \hline
\end{tabular}
\end{table}

\begin{itemize}
    \item $a$ - semi-major axis (m);
    \item $e$ - eccentricity;
    \item $i$ - inclination (rad);
    \item $\Omega$ - longitude of the ascending node (rad);
    \item $\omega$ - argument of periapsis (rad);          
    \item $v$ - mean anomaly (rad);
    \item epoch - reference epoch (mjd2000);
    \item $r$ - radius (m).
\end{itemize}

\clearpage

\section*{Appendix C: an example of maneuvers and result values}

Table \ref{tab:mans} and Table \ref{tab:agls_for_ex} show maneuvers and Environment values obtained by different agent models for the dangerous situation described in Appendix B.

\begin{table}[htbp]
\fontsize{10}{10}\selectfont
\caption{Maneuvers}
\label{tab:mans}
\centering 
\begin{tabular}{l|l|l|l|l}
\hline
                         & $dV_x$ & $dV_y$ & $dV_z$ & epoch (mjd2000) \\ \hline
baseline                 & 0.077  & 0.005  & -0.03  & 6599.962        \\ 
GS                       & 0.089  & 0.005  & -0.034 & 6599.962        \\ 
GS+CE                    & -0.072           & 0.235  & -0.098 & 6599.962        \\ \hline
CE in-track half         & -0.088 & -0.005 & 0.033  & 6599.962        \\ 
CE in-plane half         & 0.078  & -0.187 & 0.07   & 6599.962        \\ 
CE out-of-plane half        & 0.091  & 0.343  & 0.469  & 6599.962        \\ \hline
CE in-track auto         & -0.106 & -0.15  & 0.116  & 6599.950         \\ 
CE in-plane auto         & -0.058 & -0.113 & 0.079  & 6599.951        \\ 
CE out-of-plane auto        & -0.191 & -0.219 & 0.022  & 6599.951        \\ \hline
\end{tabular}
\end{table}

\begin{table}[htbp]
\fontsize{9}{9}\selectfont
\centering 
\caption{Result values}
\label{tab:agls_for_ex}
\begin{tabular}{l|l|l|l|l|l|l|l|l|l}
\hline
                  & $P_{collision}$ & Fuel  & Dev. $a$   & Dev. $e$ & Dev. $i$ & Dev. $\Omega$   & Dev. $\omega$   & Dev. $v$   & Reward   \\ \hline
threshold         & 0.0001      & 1.0   & 200.0    & 0.01   & 0.01   & 0.01     & 0.01     & -        & -7.0     \\ \hline
without maneuvers & 8.54e-03   & 0.0   & -0.0     & 0.0    & 0.0    & 0.0      & -0.0     & 0.0      & -761.0 \\ \hline
baseline          & 9.98e-05    & 0.083 & -172.241 & -1e-05 & 0.0    & 0.0      & 0.00696  & -0.00706 & -2.639   \\ 
GS                & 3.68e-05    & 0.095 & -197.142 & -1e-05 & 0.0    & 0.0      & 0.00797  & -0.00809 & -2.247   \\ 
GS+CE             & 1.17e-05    & 0.265 & 44.135   & -3e-05 & 0.0    & 0.0      & -0.00898 & 0.00894  & -1.503   \\ \hline
CE in-track half  & 4.87e-05    & 0.094 & 194.788  & 1e-05  & 0.0    & 0.0      & -0.00779 & 0.00791  & -2.335   \\ 
CE in-plane half  & 3.68e-05    & 0.215 & -81.006  & 2e-05  & 0.0    & 0.0      & 0.0089   & -0.00889 & -1.88    \\ 
CE out-of-plane half & 4.52e-05    & 0.589 & 132.594  & -0.0   & 5e-05  & -1e-4 & -0.00801 & 0.00816  & -2.521   \\ \hline
CE in-track auto  & 2e-07       & 0.217 & -127.87  & 3e-05  & 0.0    & 0.0      & 0.00092  & -0.00091 & -0.954   \\ 
CE in-plane auto  & 1.7e-06     & 0.15  & -122.371 & 2e-05  & 0.0    & -0.0     & 0.00222  & -0.00224 & -1.004   \\ 
CE out-of-plane auto & 2.3e-06     & 0.291 & -88.725  & 4e-05  & 1e-05  & 3e-05    & -0.00178 & 0.00178  & -0.943   \\ \hline
\end{tabular}
\end{table}

\begin{itemize}
    \item Rows:
    \begin{itemize}
        \item threshold -- defined threshold requirements;
        \item without maneuvers -- results of simulation without maneuvers.
    \end{itemize}
    \item Columns:
        \begin{itemize}
            \item Coll. Prob. -- total collision probability;
            \item Fuel -- fuel consumption ($m^2/s$);
            \item Dev. $a$ -- semi-major axis deviation (m);
            \item Dev. $e$ -- eccentricity deviation;
            \item Dev. $i$ -- inclination deviation (rad);
            \item Dev. $\Omega$ - longitude of the ascending node deviation (rad);           
            \item Dev. $\omega$ -- argument of periapsis deviation (rad);          
            \item Dev. $v$ -- mean anomaly deviation (rad).
        \end{itemize}
\end{itemize}

\clearpage

\section*{Appendix D: an example of conjunction}

For the example of the dangerous situation described in Appendix B, Table \ref{tab:conj_wo_for_ex} and Table \ref{tab:conj_for_ex} show information about conjunctions (miss distance $< 2000$ m) without maneuvers and Conjunctions with maneuvers obtained with the "CE out-of-plane auto" algorithm respectively.

\begin{table}[htbp]
\fontsize{10}{10}\selectfont
\centering 
\caption{Conjunctions without maneuvers}
\label{tab:conj_wo_for_ex}
\begin{tabular}{|l|l|l|l|l|l|}
\hline
   & debris name & miss distance (m) & epoch (mjd2000) & collision probability & collision danger \\ \hline
1  & DEBRIS0     & 307.033       & 6600.0          & 0.0024134             & True             \\ \hline
2  & DEBRIS6     & 226.991       & 6600.061        & 0.0019874             & True             \\ \hline
3  & DEBRIS8     & 1544.347      & 6600.238        & 0.0                   & False            \\ \hline
4  & DEBRIS1     & 750.614       & 6600.389        & 0.0001279             & True             \\ \hline
5  & DEBRIS5     & 367.326       & 6600.581        & 0.00133               & True             \\ \hline
6  & DEBRIS7     & 440.747       & 6600.597        & 0.0008903             & True             \\ \hline
7  & DEBRIS9     & 617.282       & 6600.652        & 0.0005554             & True             \\ \hline
8  & DEBRIS2     & 983.557       & 6600.791        & 1.32e-05              & False            \\ \hline
9  & DEBRIS3     & 477.438       & 6600.887        & 0.0009085             & True             \\ \hline
10 & DEBRIS4     & 617.896       & 6600.923        & 0.0003476             & True             \\ \hline
\end{tabular}
\end{table}

\begin{table}[htbp]
\fontsize{10}{10}\selectfont
\centering 
\caption{Conjunctions with maneuvers obtained with the "CE out-of-plane auto" algorithm}
\label{tab:conj_for_ex}
\begin{tabular}{|l|l|l|l|l|l|}
\hline
  & debris name & miss distance (m) & epoch (mjd2000) & collision probability & collision danger \\ \hline
1 & DEBRIS0     & 1254.839      & 6600.0          & 7e-07                 & False            \\ \hline
2 & DEBRIS6     & 1456.678      & 6600.061        & 0.0                   & False            \\ \hline
3 & DEBRIS8     & 1295.103      & 6600.238        & 1.6e-06               & False            \\ \hline
4 & DEBRIS5     & 1493.772      & 6600.957        & 0.0                   & False            \\ \hline
\end{tabular}
\end{table}

\clearpage

\bibliographystyle{AAS_publication}   
\bibliography{references}   

\end{document}